\documentclass[12pt,onecolumn,letterpaper,tightenlines,titlepage]{revtex4-1}
\usepackage[letterpaper,hmargin=1in,vmargin=1in,bmargin=1in,rmargin=1in]{geometry}
\usepackage{textcomp}
\usepackage{amsmath}
\usepackage{graphicx}
\usepackage{amssymb}
\fontfamily{times-ttf}
\begin{document}

\title{Atomistic study of the long-lived quantum coherences in the
Fenna-Matthews-Olson complex}

\author{Sangwoo Shim, Patrick Rebentrost, St\'ephanie Valleau, Al\'an Aspuru-Guzik}
\affiliation{Department of Chemistry and Chemical Biology,Harvard University, Cambridge, Massachusetts 02138, USA}

\maketitle

\section*{Abstract}
A remarkable amount of theoretical research has been carried out to
elucidate the physical origins of the recently observed long-lived
quantum coherence in the electronic energy transfer process in
biological photosynthetic systems.  Although successful in many
respects, several widely used descriptions only include an effective
treatment of the protein-chromophore interactions. In this work, 
by combining an all-atom molecular dynamics
simulation, time-dependent density functional theory, and open quantum
system approaches, we successfully simulate the dynamics of the
electronic energy transfer of the  Fenna-Matthews-Olson pigment-protein
complex. 
The resulting characteristic beating of populations and 
quantum coherences is in good agreement with the experimental results 
and the hierarchy equation of motion approach. The experimental absorption, 
linear and circular dichroism spectra
and dephasing rates are recovered at two different temperatures. 
In addition, we provide an extension of our method to
include zero-point fluctuations of the vibrational environment.  This
work thus presents one of the first steps to explain the role of
excitonic quantum coherence in photosynthetic light-harvesting
complexes based on their atomistic and molecular description.
\\
\\
\emph{Key words:} molecular dynamics, atomistic description, open quantum system, quantum chemistry, exciton dynamics

\clearpage

\section{Introduction}
\label{sec:intro}
Recent experiments suggest the existence of long-lived quantum
coherence during the electronic energy transfer process in
photosynthetic light-harvesting complexes under physiological
conditions \citep{Engel2007,Lee2007,Panitchayangkoon2010}.  This
has stimulated many researchers to seek for the physical origin of such a
phenomenon. 
The role and implication of quantum coherence during the energy transfer have
been explored in terms of the theory of open quantum systems
\citep{Rebentrost2009b,Plenio2008,Jang2008,Palmieri2009,Ishizaki2009a,Ishizaki2009b,Ishizaki2009,Rebentrost2009,Strumpfer2009,Virshup2009,Caruso2009,Wu2010,Dijkstra2010},
and also in the context of quantum information and entanglement
\citep{Caruso2010a,Fassioli2010,Sarovar2010,Rebentrost2011}.
However, the characteristics of the protein environment, and especially 
its thermal vibrations or phonons, have not been fully investigated from the 
molecular viewpoint. A more detailed description of the bath in atomic 
detail is desirable; to investigate the structure-function relationship 
of the protein complex and to go beyond the assumptions used in popular
models of photosynthetic systems.

Protein complexes constitute one of the most essential components in
every biological organism. They remain one of the major targets of
biophysical research due to their tremendously diverse and, in some
cases, still unidentified structure-function relationship.  Many
biological units have been optimized through evolution and the
presence of certain amino acids rather than others is fundamental for
functionality~\citep{Tamoi2010,Jahns2002,Pesaresi1997}.
In photosynthesis, one of the most well-characterized pigment-protein complexes is the 
Fenna-Matthews-Olson (FMO) complex which is a light-harvesting complex found in green
sulphur bacteria. It functions as an intermediate conductor for exciton transport 
located between the antenna complex where light is initially absorbed and the reaction
center.  Since the resolution of its crystal structure over 30 years
ago \cite{Olson2004}, the FMO trimer, composed of 3 units each comprising 
8 bacteriochlorophylls has been extensively studied both experimentally 
\cite{Freiberg1997, Savikhin1998,Wendling2000,Tronrud2009} and theoretically 
\cite{Vulto1998,Adolphs2006}. For instance regarding the structure-function relationship, 
it has been shown~\citep{Renger2011} that
amino acid residues cause considerable shifts in the site energies 
of bacteriochlorophyll $a$ (BChl) molecules of the FMO complex 
and in turn causes changes to the energy transfer properties.

Have photosynthetic systems adopted interesting quantum effects to improve their
efficiency in the course of evolution, as suggested by the experiments? 
In this article, we provide a first step to answer this
question by characterizing the protein environment of the FMO photosynthetic system
to identify the microscopic origin of the long-lived quantum coherence.
We investigate the quantum energy transfer of a
molecular excitation (exciton) by incorporating an all-atom molecular
dynamics (MD) simulation. The molecular energies are computed with
time-dependent density functional theory (TDDFT) along the MD
trajectory.  The evolution of the excitonic density matrix is obtained
as a statistical ensemble of unitary evolutions by a time-dependent Schr\"odinger equation. 
Thus, this work is in contrast to many studies based on quantum master equations in
that it includes atomistic detail of the protein environment into the
dynamical description of the exciton.  We also introduce a novel
approach to add quantum corrections to the dynamics. 
Furthermore, a quantitative comparison to the hierarchical
equation of motion and the Haken-Strobl-Reineker method is presented.  As the main result, the time
evolution of coherences and populations shows characteristic beatings
on the time scale of the experiments. Surprisingly, we observe that
the cross-correlation of site energies does not play a significant
role in the energy transfer dynamics.

The paper is structured as follows: In the first part we present the
methods employed and in the second part the results followed by
conclusions.  In particular, the partitioning of the system and bath
Hamiltonian in classical and quantum degrees of freedom and details of
the MD simulations and calculation of site energies are discussed in
Section \ref{subsec:md}.  The exciton dynamics of the system under the
bath fluctuations is then presented in Section
\ref{subsec:exciton}. In Section \ref{subsec:mctdjump} we introduce
a quantum correction to the previous exciton dynamics. 
Using the discussed methods we evaluated site
energies and their distribution at 77 and 300K in Section
\ref{subsec:energies} and we also computed the linear absorption
spectrum of the FMO complex in Section \ref{subsec:spectrum}.  The
site basis dephasing rates are discussed in Section
\ref{subsec:dephasing_rates}.  From the exciton dynamics of the system
we obtained populations and coherences and compared to the QJC-MD
approach in Section \ref{subsec:populations}.  We then compare the
MD and quantum corrected MD methods to the hierarchical equation of motion
(HEOM) and Haken-Strobl-Reineker (HSR) methods in Section
\ref{subsec:comparison}.
In Section \ref{subsec:corr} we determined the spectral density for
each site from the energy time bath-correlator and studied the effect
of auto and cross-correlations on the exciton dynamics by introducing
a comparison to first-order autoregressive processes.  We conclude in
Section \ref{sec:conclusion} by summarizing our results.

\section{Methods}
\label{sec:methods}

\subsection{Molecular Dynamics Simulations}
\label{subsec:md}
A computer simulation of the quantum evolution of the entire FMO complex is 
certainly unfeasible with the
currently available computational resources. However, we are only interested
in the electronic energy transfer dynamics among BChl
molecules embedded in the protein support.
This suggests a decomposition of the total system Hamiltonian operator into three parts:
the relevant system, the bath of vibrational modes, and the system-bath interaction Hamiltonians.
The system Hamiltonian operates on the excitonic system alone which is defined by a set of 
two-level systems. Each two-level system represents the ground and first excited electronic state 
of a BChl molecule. 
In addition, the quantum mechanical state of the exciton is assumed to be restricted to the
single-exciton manifold because the exciton density is low. 
On the other hand, factors affecting the system site
energies have intractably large degrees of freedom, so it is reasonable to
treat all those degrees of freedom as the bath of an open quantum system.

More formally, to describe the system-bath interplay by including atomistic detail of the bath,
we start from the total Hamiltonian operator and decompose it
in a general way such that no assumptions on the functional
form of the system-bath Hamiltonian are necessary \cite{May2004}:
\begin{eqnarray}\label{eq:Htotal}
  {\hat{H}}_{total} &=& \sum_{m} \int d\boldsymbol{R}\ \epsilon_m (\boldsymbol{R}) |m\rangle \langle m| \otimes |\boldsymbol{R}\rangle \langle \boldsymbol{R}| \nonumber\\
  &&+ \sum_{m,n} \int d\boldsymbol{R}\ \left\{ J_{mn} (\boldsymbol{R}) |m\rangle \langle n| \otimes |\boldsymbol{R}\rangle \langle \boldsymbol{R}| + c.c. \right\} \nonumber\\
  &&+ |\mathbf{1}\rangle \langle \mathbf{1}| \otimes \hat{T}_{\boldsymbol{R}} + \sum_{m} \int d\boldsymbol{R}\ V_m(\boldsymbol{R}) |m \rangle \langle m| \otimes |\boldsymbol{R}\rangle \langle \boldsymbol{R}|.
\end{eqnarray}
Here,
$\boldsymbol{R}$ corresponds to the nuclear coordinates of the FMO complex
including both BChl molecules, protein, and enclosing water molecules.
The set of states $|m\rangle \otimes |\boldsymbol{R}\rangle$ denote the 
presences of the exciton at site $m$ given that the FMO complex is in the 
configuration $\boldsymbol{R}$, $\epsilon_m(\boldsymbol{R})$ represents 
the site energy of the $m$th site and $J_{mn}(\boldsymbol{R})$ is
the coupling constant between the $m$th and $n$th sites.
Note that the site energies and coupling terms can be modulated by
$\boldsymbol{R}$.
$|\mathbf{1}\rangle \langle\mathbf{1}|$ is the identity operator
in the excitonic subspace,
$\hat{T}_{\boldsymbol{R}}$ is the kinetic operator for the nuclear coordinates
of the FMO complex, and
${V_m(\boldsymbol{R})}$ is the potential energy
surface for the complex when the exciton at site $m$ under
Born-Oppenheimer approximation.
Given multiple Born-Oppenheimer surfaces, one would need to carry out a
coupled nonadiabatic propagation. However, as a first approximation,
we assume that the
change of Born-Oppenheimer surfaces does not affect the bath dynamics
significantly. This approximation becomes better at small reorganization energies.
Indeed, BChl molecules have significantly smaller reorganization energies 
than other chromophores \cite{Mckenzie2008}.
With this assumption, we can ignore the dependence on
the excitonic state in the $V$ term, thus the system-bath
Hamiltonian only contains the one-way influence from the bath to the
system. We also adopted Condon approximation so that the J terms do not depend
on $\boldsymbol{R}$:
\begin{eqnarray}\label{eq:HtotalApprox}
  H_{S} &=& \sum_{m} \int d\boldsymbol{R}\ \bar{\epsilon}_m |m \rangle \langle m| \otimes |\boldsymbol{R}\rangle \langle \boldsymbol{R}|
  + \sum_{m,n} \int d\boldsymbol{R}\ \left\{ J_{mn}(\boldsymbol{R}) |m \rangle \langle n| \otimes |\boldsymbol{R}\rangle \langle \boldsymbol{R}| + c.c. \right\},\nonumber\\
  &\approx& \sum_{m} \int d\boldsymbol{R}\ \bar{\epsilon}_m |m \rangle \langle m| \otimes |\boldsymbol{R}\rangle \langle \boldsymbol{R}|
  + \sum_{m,n} \int d\boldsymbol{R}\ \left\{ \bar{J}_{mn} |m \rangle \langle n| \otimes |\boldsymbol{R}\rangle \langle \boldsymbol{R}| + c.c. \right\},\nonumber\\
  H_{B} &=& |\mathbf{1}\rangle \langle\mathbf{1}|\otimes \hat{T}_{\boldsymbol{R}}
  + \sum_{m} \int d\boldsymbol{R}\ {V}_m(\boldsymbol{R}) |m\rangle \langle m| \otimes |\boldsymbol{R}\rangle \langle \boldsymbol{R}|,\nonumber\\
  &\approx& |\mathbf{1}\rangle \langle \mathbf{1}| \otimes \hat{T}_{\boldsymbol{R}}
  +  \int d\boldsymbol{R}\ {V_{ground}}(\boldsymbol{R}) |\mathbf{1}\rangle \langle \mathbf{1}| \otimes |\boldsymbol{R}\rangle \langle \boldsymbol{R}|, \nonumber\\
  H_{SB} &=& \sum_{m} \int d\boldsymbol{R}\ \left\{\epsilon_m(\boldsymbol{R})
  - \bar{\epsilon}_m \right\} |m\rangle \langle m| \otimes |\boldsymbol{R}\rangle \langle \boldsymbol{R}|,\nonumber\\
  {H}_{total} &=& H_{S} + H_{B} + H_{SB}.
\end{eqnarray}
Based on this decomposition of the total Hamiltonian, we set up a
model of the FMO complex with the AMBER 99 force
field~\citep{Cornell1995,Ceccarelli2003} and approximate the
dynamics of the protein complex bath by classical mechanics.
The initial configuration of the MD simulation was taken
from the x-ray crystal structure of the FMO
complex of \emph{Prosthecochloris aestuarii} (PDB ID: 3EOJ.). 
Shake constraints were used for all bonds containing hydrogen 
and the cutoff distance for the long range interaction was chosen to be 12 \AA. 
After a 2ns long equilibration run, the production run was obtained for a total
time of 40ps with a 2fs timestep. For the calculation of the optical gap, 
snapshots were taken every 4fs.
Two separate simulations at 77K and 300K were carried out with an
isothermal-isobaric (NPT) ensemble to investigate the temperature dependence of
the bath environment. Then, parameters for the system and the system-bath
Hamiltonian were calculated using quantum chemistry methods along the
trajectory obtained from the MD simulations.

We chose not to include the newly resolved eighth BChl molecule
\cite{Renger2011} in our simulations because up to now, the large majority 
of the scientific community has focused on the seven site system which is therefore 
a better benchmark to compare our calculations to previous work. It is important to 
note however that this eighth site may have an important role on the dynamics. 
In particular, as suggested in \cite{Ritschel2011, Moix2011} this eighth site is
considered to be the primary entering point for the exciton in the FMO complex 
and its position dictates a preferential exciton transport pathway rather than 
two independent ones. Also when starting with an exciton on this eighth site, 
the oscillations in the coherences are largely suppressed. 

The time-dependent site energy $\epsilon_m$ was evaluated as the excitation
energy of the $Q_y$ transition of the corresponding
BChl molecule.
We employed the time-dependent density functional theory (TDDFT) with BLYP
functional within the Tamm-Dancoff approximation (TDA) using the Q-Chem quantum
chemistry package~\citep{Shao2006}.
The basis set was chosen to be 3-21G after considering
a trade-off between accuracy and computational cost.
The $Q_y$ transition was identified as the excitation with the highest
oscillator strength among the first 10 singlet excited states. Then, the
transition dipole of the selected state was verified to be parallel to the
$y$ molecular axis.
Every atom which did not belong to the TDDFT target molecule was incorporated
as a classical point charge to generate the
external electric field for the QM/MM calculation.
Given that the separation between BChl molecules 
and the protein matrix is quite clear, employing 
this simple QM/MM method with classical external charges 
to calculate the site energies is a good approximation.  
The external charges were taken from the partial charges of the AMBER force
field~\citep{Cornell1995,Ceccarelli2003}. The coupling terms, $J_{mn}$, can also
be obtained from quantum chemical approaches like transition density cube or 
fragment-excitation difference methods~\citep{Krueger1998,Hsu2008}. However,
in this case we employed the MEAD values of the couplings of the Hamiltonian 
presented in the literature~\citep{Adolphs2006} and considered them to be constant in
time. $\bar{\epsilon}_m$ was straightforwardly chosen as the time averaged site
energy for the $m$th site.

\subsection{Exciton Dynamics}
\label{subsec:exciton}
In this section, we describe the method for the dynamics of the excitonic reduced 
density matrix within our molecular dynamic simulation framework. 
It is based on a simplified version of the quantum-classical hybrid method (Ehrenfest) described in \cite{May2004}. 
The additional assumption on Hamiltonian (\ref{eq:HtotalApprox}) is that the bath 
coordinate $\boldsymbol{R}$ is a classical variable, denoted by a superscript ``cl". As 
discussed above, the time-dependence of these variables arises from the Newtonian MD 
simulations. The additional force on the nuclei due to the electron-phonon coupling \cite{May2004} 
is neglected.
Hence, the Schr\"odinger equation for the excitonic system is given by:
\begin{equation}\label{eq:QuantumClassical}
i\hbar \frac{\partial}{\partial t} |\psi(t)\rangle \approx \left\{ H_S+
H_{SB}(\boldsymbol{R}^{cl}(t)) \right\} |\psi(t)\rangle.
\end{equation}
The system-environment coupling leads to an effective time-dependent Hamiltonian
$H_{eff}(t) = H_S + H_{SB}(\boldsymbol{R}^{cl}(t))$.
This equation suggests a way to propagate the reduced
density matrix as an average of unitary evolutions given 
by Eq. (\ref{eq:QuantumClassical}). 
First, short MD trajectories (in our case $1$ ps long) are uniformly sampled 
from the full MD trajectory ($40$ ps). 
Then, for each short MD trajectory, the excitonic system can be propagated under 
unitary evolution with a simple time-discretized exponential integrator.
The density matrix is the classical average of these unitary evolutions:
\begin{eqnarray}
  \rho_S(t) &=& \frac{1}{M}\sum_{i=1}^M |\psi_i(t)\rangle \langle \psi_i(t)|,
\end{eqnarray}
where $M$ is the number of sample short trajectories.
Each trajectory is subject to different time-dependent fluctuations from the
bath, which manifests itself as decoherence when averaged to the statistical
ensemble. Compared to many methods based on the stochastic unraveling of the
master equation, e.g.~\citep{Dalibard1992,Piilo2008},
our formalism directly utilizes the fluctuations generated by the MD simulation.
Therefore, the detailed interaction between system and bath is captured.  
The temperature of the bath is set by the thermostat of the MD simulation, thus
no further explicit temperature dependence is required in the overall dynamics.
The dynamics obtained by this numerical integration of the Schr\"odinger equation
will also be compared to the HEOM approach. The HEOM is
briefly described in the Supporting Material along with a discussion on the differences
respect to the MD-method.

\subsection{Quantum Jump Correction to MD Method (QJC-MD)}
\label{subsec:mctdjump}
The MD/TDDFT simulation above leads to
crucial insights into the exciton dynamics. However, it does not
capture quantum properties of the vibrational environment such as
zero-point fluctuations.  At zero temperature all the atoms in the MD
simulation are completely frozen. Moreover, similarly to an
infinite-temperature model, at long times of the quantum dynamical
simulation the exciton is evenly distributed among all molecules, as we will see below. 
In order to obtain a more realistic description, we modify the
stochastic simulation by introducing quantum jumps derived from the
zero-point (zp) fluctuations of the modes in the vibrational
environment. We refer to this corrected version of the MD propagation
as QJC-MD.

Introducing harmonic bath modes explicitly we reformulate the system-bath Hamiltonian as:
\begin{equation}
H_{SB} = \sum_{m} |m\rangle \langle m| \sum_{\xi} g_{\xi}^{m} R_{\xi}.
\end{equation}
Here, each $g_{\xi}^{m}$ represents the coupling strength of a site $m$ to a particular mode $\xi$ and
$R_{\xi}$ is the dimensionless position operator for that mode. 
We now formulate our correction by separating the bath operators into two parts, 
$R_{\xi}=R_{\xi}^{zp}+R_{\xi}^{MD}$, 
the first part is due to zero-point fluctuations and the second 
comes from our MD simulations. As above, the MD part is replaced by the classical time-dependent variables, $R_{\xi}^{MD} \to R_{\xi}^{cl}(t)$.
The zero-point operator is expressed by creation and annihilation operators, 
$R_{\xi}^{zp}=b_{\xi}^{zp}+b_{\xi}^{zp,\dagger}$, which satisfy the usual commutation 
relations $[b_{\xi}^{zp},b_{\xi\prime}^{zp,\dagger}]=\delta_{\xi \xi\prime}$. 
By construction, for the zp-fluctuations one has 
$\langle b_{\xi}^{zp,\dagger}b_{\xi}^{zp} \rangle = 0$.

The zp-fluctuations can only induce excitonic transitions from higher to
lower exciton states in the instantaneous eigenbasis of the
Hamiltonian, thus leading to relaxation of the excitonic system. The
evolution of the populations $P_{M}$ of the instantaneous eigenstates
$|M\rangle (t)$ due to the zero-point correction is expressed by a Pauli master equation as:
\begin{eqnarray}\label{eqTDJPopulations}
\left(\dot{P}_{M}\right)_{zpc} &=&-\sum_{N} \gamma(\omega_{MN}) P_{M} + \sum_{N} \gamma(\omega_{NM}) P_{N},
\end{eqnarray}
and for the coherences as:
\begin{equation}\label{eqTDJCoherences}
\left( \dot{C}_{MN}\right)_{zpc}=-\frac{1}{2} \gamma(|\omega_{MN}|) C_{MN}.
\end{equation}
The associated rate can be derived from a secular Markovian Redfield theory \cite{Breuer2002}
to be $\gamma(\omega_{MN}) =2\pi J\left( \omega _{MN}\right) \sum_{m}|c_{m}\left(
M\right) |^2|c_{m}\left( N\right) |^2$, where the
spectral density $J\left( \omega \right)$ is only non-zero for positive
transition frequencies $\omega_{MN}=E_M-E_N$ and taken to be as in ~\citep{Ishizaki2009a}.
The coefficients $c_{m}\left( M\right) $ translate from site to energy
basis. The time evolution given by Equations (\ref{eqTDJPopulations}) and (\ref{eqTDJCoherences})
is included in the dynamics simulation by introducing quantum jumps as in the 
Monte-Carlo wavefunction (MCWF) method \cite{Dalibard1992}. 
We thus arrive at a hybrid classically averaged $H\left( t\right)$ simulation with additional quantum
transitions induced by the vacuum fluctuations of the vibrational modes.

\section{Results and Discussion}
\label{sec:result}

\subsection{Site energy distributions}
\label{subsec:energies}

\begin{figure}
  \includegraphics[width=3.25in,keepaspectratio]{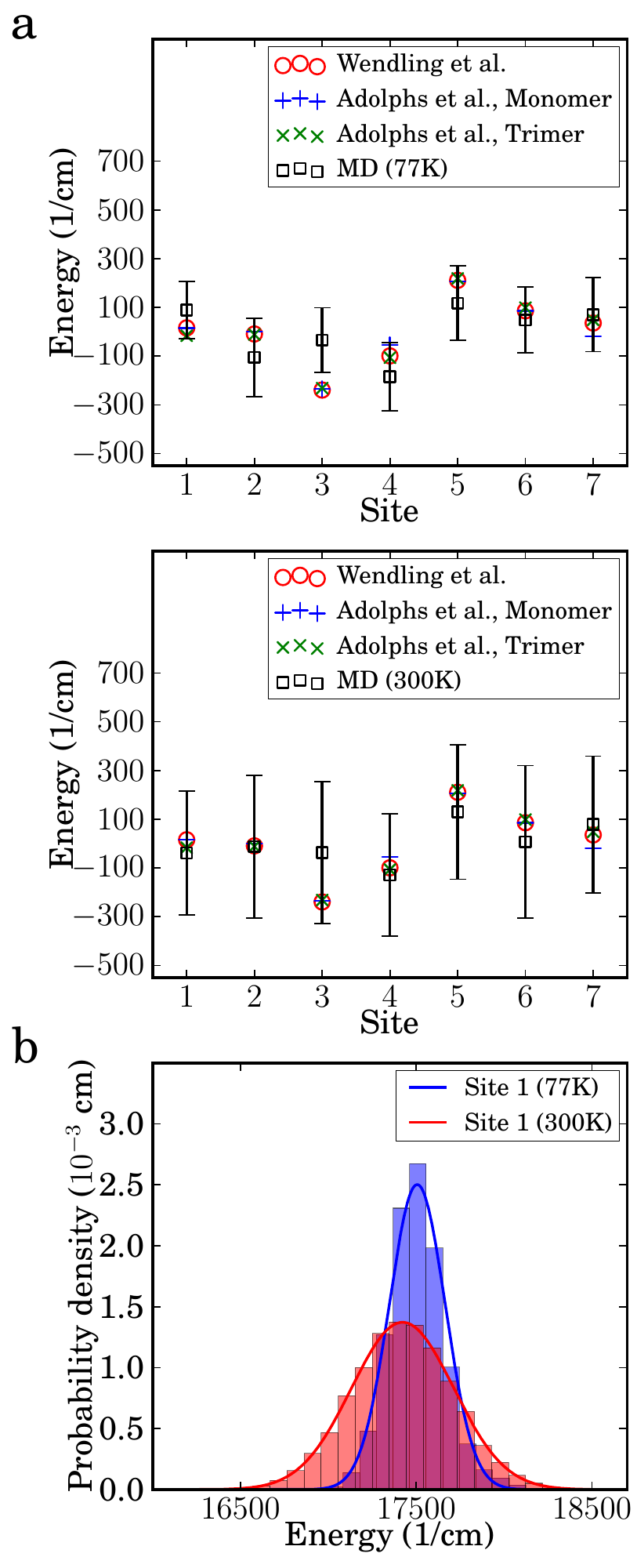}
  \caption{}
  \label{fig:siteenergy_distribution}
\end{figure}

Using the coupled QM/MD simulations, site energies were obtained for 
each BChl molecule. These energies and their fluctuations are 
reported in Figure \ref{fig:siteenergy_distribution}.  
We note that the magnitude of the fluctuations are of the order of hundreds of cm$^{-1}$.  
Although the order of the site
energies does not perfectly match previously reported
results~\citep{Wendling2002,Adolphs2006}, the overall trend does not deviate much,
especially considering that our result is purely based on \emph{ab
initio} calculations without fitting to the experimental result.
The $Q_y$ transition energies calculated by TDDFT are known to be
systematically blue-shifted with respect to the experiment~\cite{Vokacova2007}.
However, the scale of the fluctuations remains reasonable. Therefore,
the comparison in Fig. \ref{fig:siteenergy_distribution} was made after shifting
the overall mean energy to zero for each method.

The excitation energy using TDDFT does not always converge when the
configuration of the molecule deviates significantly from its ground
state structure. The number of points which failed to converge was on average
less than 4\% for configurations at 300K, and less than 2\% at 77K.
We interpolated the original time series to obtain smaller time steps 
and recover the missing points. Interpolation could lead to severe distortion 
of the marginal distribution when the number of available points is too small. 
However, in our case, the distributions virtually remained the same with 
and without interpolation.

\subsection{Dephasing rates}
\label{subsec:dephasing_rates}
\begin{figure}
  \includegraphics[width=3.25in,keepaspectratio]{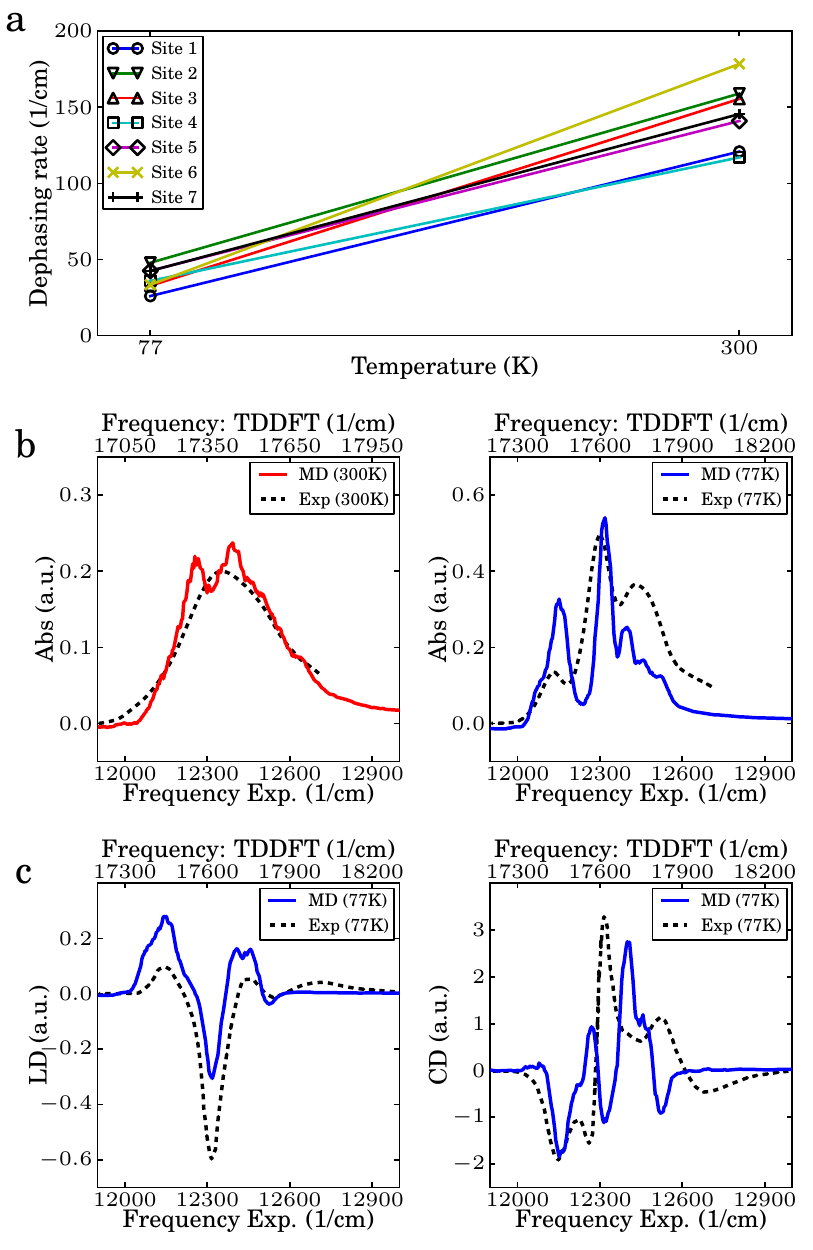}
  \caption[format=plain,justification=centerlast]{}
  \label{fig:dprates_spectra}
\end{figure}

  In the Markovian approximation and assuming an exponentially
decaying autocorrelation function, the dephasing rate
$\gamma_\phi$ is proportional to the variance of the site
energy $\sigma_\epsilon^2$ ~\citep{Breuer2002}:
\begin{eqnarray} \label{eqDephasingRate}
  \gamma_\phi = \frac{2}{\hbar} \sigma_\epsilon^2\tau,
\end{eqnarray}
where $\tau$ is a time decay parameter which we estimated through a comparison
to first order autoregressive processes, as described in Section \ref{subsec:corr}.
The dependence on the variance is clearly justified: states associated with large 
site energy fluctuations tend to undergo faster dephasing. 
Figure \ref{fig:dprates_spectra}, panel a), presents the
approximate site basis dephasing rates for each site with $\tau \approx$ 5fs.
The averaged value of the slopes is 0.485 cm$^{-1}$ K$^{-1}$, which is
in good agreement with the experimentally measured value of
0.52 cm$^{-1}$ K$^{-1}$ obtained from a closely related species
\emph{Chlorobium tepidum}
in the exciton basis~\citep{Panitchayangkoon2010}.
From this plot we note the presence of 
a positive correlation between temperature and dephasing rate. 
This correlation is plausible: as temperature increases so does the energy disorder, hence 
the coherences should decay faster. In fact, in the Markovian approximation, 
dephasing rates increase linearly with temperature~\cite{Leggett1987,Breuer2002}.
Calculations at other temperatures are underway to verify this and to obtain more 
information on the precise temperature dependence of the dephasing rates.

\subsection{Simulated Spectra}
\label{subsec:spectrum}
The absorption, linear dichroism (LD), and circular dichroism (CD) spectra can
be obtained from the Fourier transform of the corresponding response functions.
The spectra can be evaluated for the seven BChl
molecules using the following
expressions~\citep{Pearlstein1991, Damjanovic2002}:
\begin{widetext}
\begin{eqnarray}\label{eqn:absorption}
  I_{Abs} (\omega) &\propto& \mbox{Re} \int_0^\infty dt \ e^{i\omega t} \sum_{m,n=1}^7
  \langle \vec{d}_m \cdot \vec{d}_n \rangle
  \{ \langle U_{mn}(t,0) \rangle - \langle U_{mn}^*(t,0) \rangle \},\nonumber\\
  I_{LD} (\omega) &\propto& \mbox{Re} \int_0^\infty dt \ e^{i\omega t} \sum_{m,n=1}^7
  \langle 3(\vec{d}_m \cdot \hat{r})(\vec{d}_n \cdot \hat{r}) -
  \vec{d}_m \cdot \vec{d}_n \rangle
  \{ \langle U_{mn}(t,0) \rangle - \langle U_{mn}^*(t,0) \rangle \},\nonumber\\
  I_{CD} (\omega) &\propto& \mbox{Re} \int_0^\infty dt \ e^{i\omega t} \sum_{m,n=1}^7
  \langle \bar{\epsilon}_{m} (\vec{R}_m - \vec{R}_n) \cdot (\vec{d}_m \times \vec{d}_n) \rangle
  \{ \langle U_{mn}(t,0) \rangle - \langle U_{mn}^*(t,0) \rangle \},
\end{eqnarray}
\end{widetext}
where $m$ and $n$ are indices for the BChl molecules in the complex,
$\vec{d}_m$ is the transition dipole moment of the $m$th site,
$U_{mn}(t, 0)$ is the $(m,n)$ element of the propagator in the site
basis, $\hat{r}$ is the unit vector in the direction of the rotational
symmetry axis, $\vec{R}_m$ is the coordinate vector of the
site $m$, and $\langle \cdots \rangle$ indicates an ensemble average.  
The ensemble average was evaluated by sampling and averaging over 4000 trajectories. 
We applied a low-pass filter to smooth out the noise originated from truncating 
the integration and due to the finite number of trajectories.
Figure \ref{fig:dprates_spectra} panel b) and c) show 
direct comparison of the calculated and experimental spectra 
at 77K and 300K. As discussed in Section \ref{subsec:energies}, 
TDDFT tends to systematically overestimate the excitation energy of 
the Q$_y$ transition~\citep{Olbrich2010} yet the fluctuation widths 
of the site energies are reasonable.  In fact, the width and overall 
shape of the calculated spectrum is in good agreement 
with the experimental spectrum at each temperature.
Calculated LD and CD spectra also reproduce well the experimental measurements,
considering that no calibration to experiments was carried out.
Since both LD and CD spectra are sensitive to the molecular structure
it appears that our microscopic model correctly captures these details.

\subsection{Population dynamics and long-lived quantum coherence}
\label{subsec:populations}

\begin{figure}
  \includegraphics[width=6.5in,keepaspectratio]{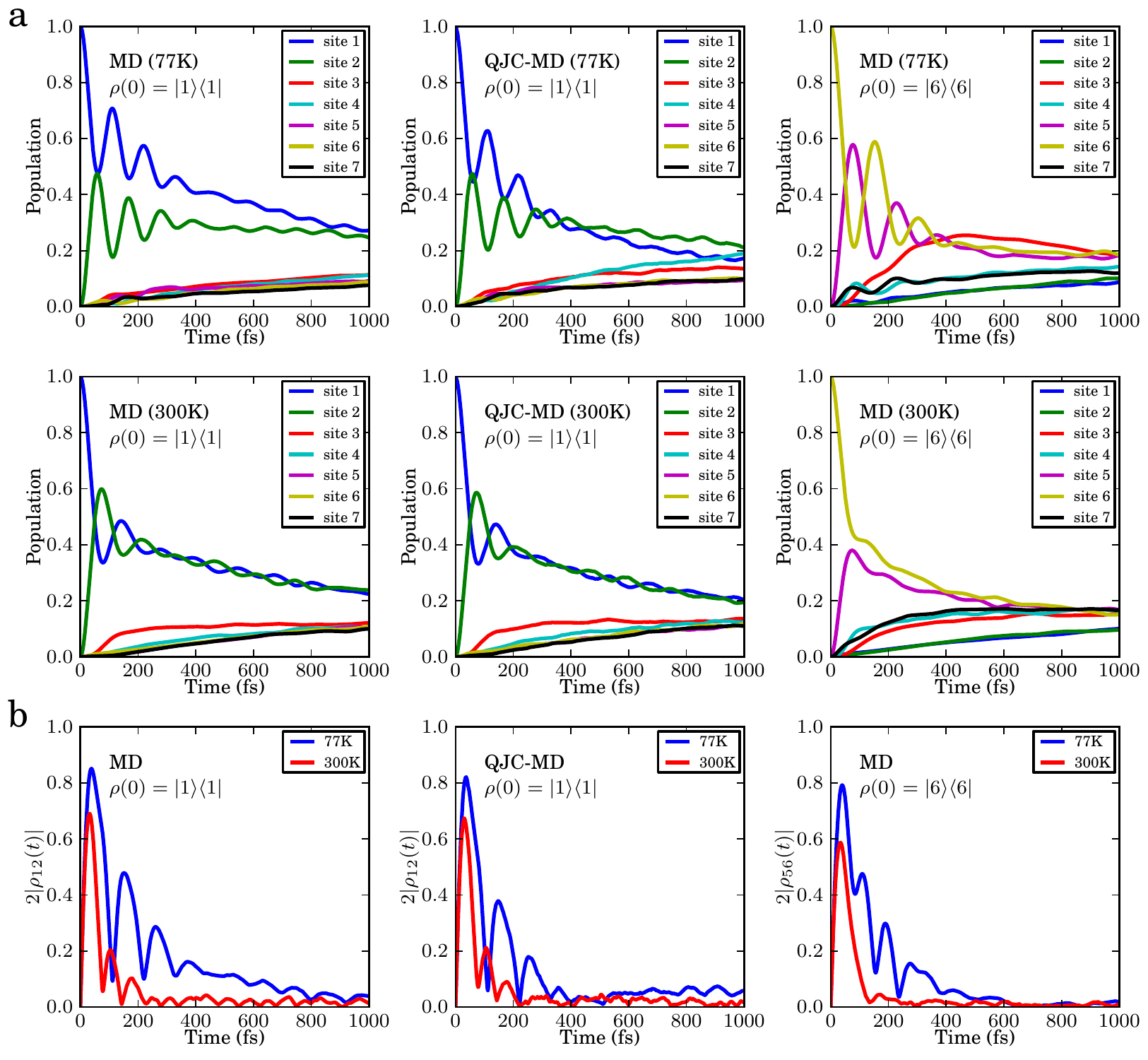}
  \caption{ 
  }
  \label{fig:population}
\end{figure}

The MD method is based on minimal assumptions and directly
evaluates the dynamics of the reduced density matrix from the total
density matrix as described in Section \ref{sec:methods}.
The reduced density matrix was obtained after averaging over
4000 trajectories. Figure \ref{fig:population} shows the 
population and coherence dynamics of each of the seven sites 
according to the dephasing induced by the nuclear motion of the FMO complex.
In particular, the populations and the absolute value of the 
pairwise coherences, as defined in ~\citep{Sarovar2010} 
($ 2\cdot \left| \rho_{12} \left(t\right) \right|$ and 
$ 2\cdot \left| \rho_{56} \left(t\right) \right|$) 
are plotted at both 77 and 300K starting with an initial state in site 1
(first three panels) and then in site 6 (last three panels).
Until very recently ~\cite{Ritschel2011,Moix2011} site 1 and 6 have been thought as 
the entry point of an exciton in the FMO complex, therefore
most of the previous literature chose the initial reduced density matrix
to be either pure states $|1\rangle\langle 1|$ or $|6\rangle \langle 6|$
\cite{Ishizaki2009,Huo2010,Wu2010}. However, our method could be applied to
any mixed initial state without modification.
We note that coherent beatings last for about 400fs at 77K and 200fs at 300K.
These timescales are in agreement with those reported for FMO 
\cite{Ishizaki2009,Panitchayangkoon2010} and with what was found in Section 
\ref{subsec:dephasing_rates}. Although quite accurate in the short time limit, 
the MD method populations do not reach thermal equilibrium at long times. 
This was verified by propagating the dynamics to twice the time shown in Figure
\ref{fig:population}. This final classical equal distribution is similar to the HSR
model result.  
The three central panels of Figure \ref{fig:population} show the same
populations and coherences obtained from the QJC-MD method.
As discussed in Section \ref{sec:methods}, this method includes a zero
point correction through relaxation transitions and predicts a more
realistic thermal distribution at 77K. At 300K the quantum
correction is less important in the dynamics because the Hamiltonian
fluctuations dominate over the zero temperature quantum fluctuations.

\subsection{Comparison between MD, QJC-MD, HEOM, and HSR methods}
\label{subsec:comparison}

\begin{figure}
  \includegraphics[width=3.25in,keepaspectratio]{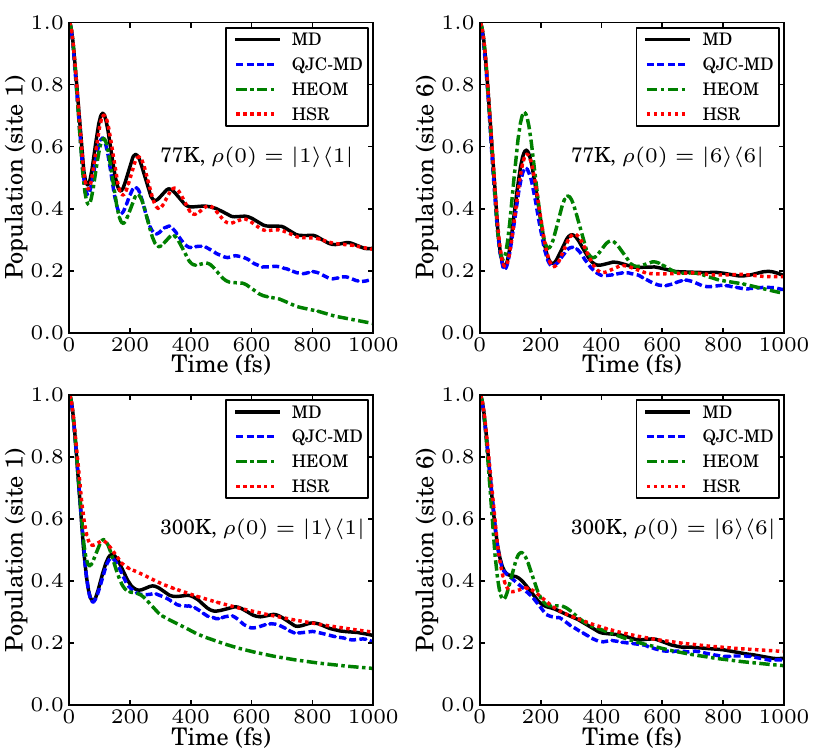}
  \caption{}
  \label{fig:populations_comparison}
\end{figure}

Figure \ref{fig:populations_comparison} shows a direct comparison of the
population dynamics of site 1 calculated using the HEOM method discussed 
by Ishizaki and Fleming~\citep{Ishizaki2009,Zhu2011}, our MD and quantum 
corrected methods at 77K and 300K, and the HSR model \cite{Haken1972,Haken1973} 
with dephasing rates obtained from Eq. (\ref{eqDephasingRate}). 
We observe that the short-time dynamics and dephasing
characteristics are surprisingly similar, considering that the methods
originate from very different assumptions. Atomistic detail can allow
for differentiation of the system-environment coupling for different
chromophores. For example, at both temperatures (right panels),
the MD populations of site 6 undergo faster decoherence than
the corresponding HEOM results. We attribute this to the difference in
energy gap fluctuations of site energy between site 1 and 6 obtained 
from the MD simulation as can be seen in \ref{fig:siteenergy_distribution}. 
On one hand, in the HEOM method, site energy fluctuations are considered 
to be identical across all sites, on the other, in our method the fluctuations 
of each site are obtained from the MD simulation in which each site is 
associated with a different chromophore-protein coupling.
Nevertheless, the fact that we obtain qualitatively similar results to 
the HEOM approach (at least when starting in 
$\rho(0)=\left|1\right\rangle\left\langle 1\right|$) without considering 
non equilibrium reorganization processes suggests that such processes 
might not be dominant in the FMO. The quantum correction results (QJC-MD), 
for every temperature and initial state, are in between the HEOM and MD results. 
This is due to the induced relaxation from zero-point fluctuations
of the bath environment, which are not included in the MD method but 
included in the QJC-MD and HEOM methods.
  
The HSR results take into account the site-dependence of the dephasing rates based on Eq. 
(\ref{eqDephasingRate}). The method is briefly described in the supplementary material. 
Due to the Markovian assumption, this model shows slightly less coherence than the HEOM method 
and similarly to the MD method it converges to an equal classical mixture of all sites in the 
long time limit.

\subsection{Correlation functions and spectral density}
\label{subsec:corr}
\begin{figure}
  \includegraphics[width=6.5in,keepaspectratio]{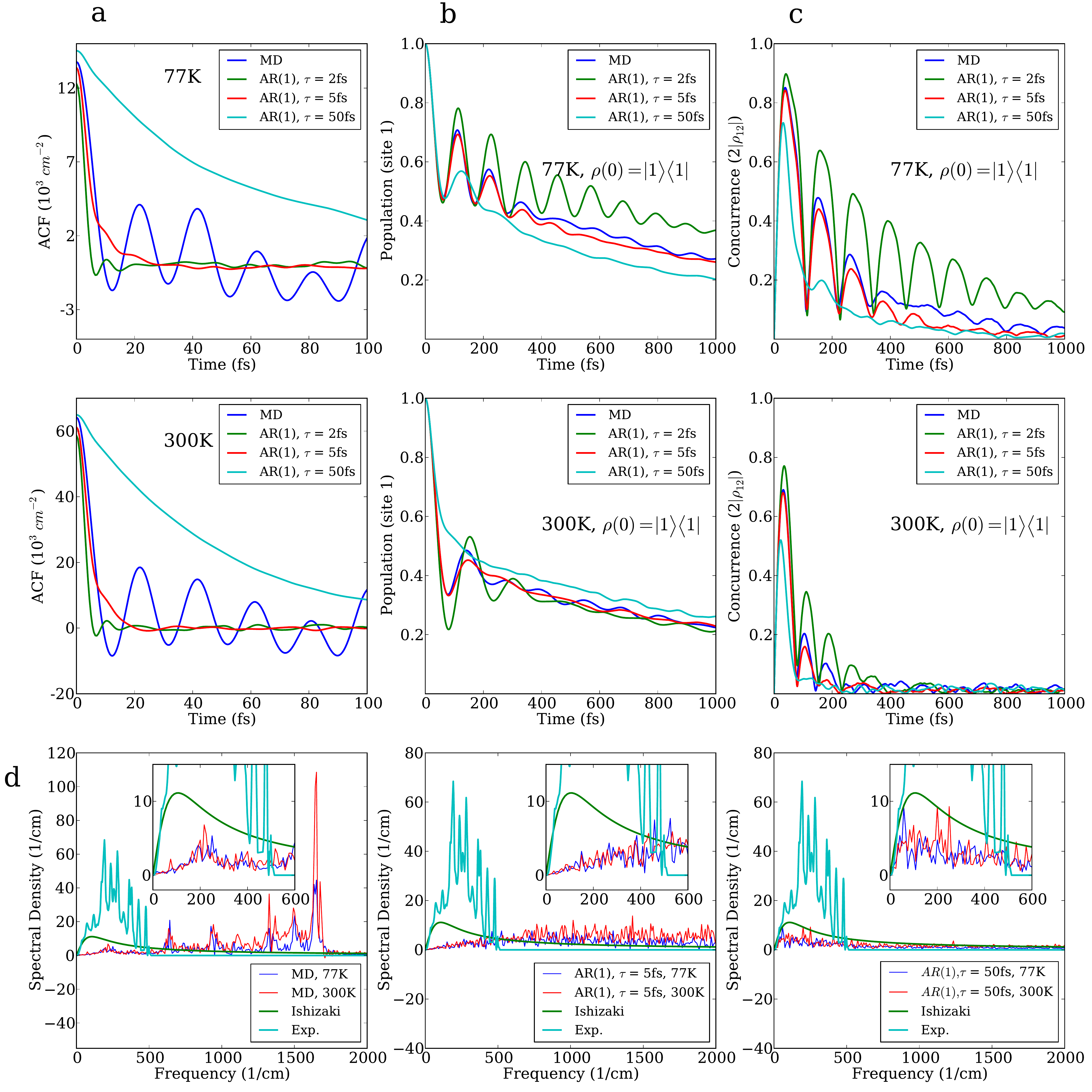}
  \caption{}
  \label{fig:autocorr_spectraldensity}
\end{figure}

The bath autocorrelation function and its spectral density contain information 
on interactions between the excitonic system and the bath. 
The bath correlation function is defined as 
$C(t) = \langle \delta \epsilon (t) \delta \epsilon(0) \rangle$ with $\delta \epsilon = \epsilon(t) - \bar{\epsilon}$. 
For the MD simulation, $C(t)$ is shown in Fig. \ref{fig:autocorr_spectraldensity} a) 
for the two temperatures. 

To study the effect of the decay rate of the autocorrelation function
on the population dynamics, we modeled site energies using first-order
autoregressive (AR(1)) processes~\citep{Percival1993}.
The marginal distribution of each process
was tuned to have the same mean and variance as for the MD simulation.
The autocorrelation function of the AR(1) process is an exponentially decaying
function:
\begin{eqnarray}
  C(t) \propto \exp(-t/\tau).
\end{eqnarray}
We generated three AR(1) processes with different time constants $\tau$
and propagated the reduced density matrix using the Hamiltonian
corresponding to each process. As can be seen in Fig.
\ref{fig:autocorr_spectraldensity}, panel a), the autocorrelation function of the
AR(1) process with $\tau \approx$ 5fs has a similar initial decay rate to
that of the MD simulation at both temperatures. Therefore, as shown in the
last three horizontal panels, its spectral density is in good agreement
with the MD simulation result in the low frequency region, i.e up to 600cm$^{-1}$.
Modes in this region are known to be the most important in the dynamics and in 
determining the the decoherence rate. Also, as panels b) and c) show, 
that same AR(1) process with $\tau \approx$ 5fs exhibits similar population 
beatings and concurrences to those of the MD simulation.
The relation of this 5fs time scale to others reported in~\cite{Ishizaki2009,Cho2005}
is presently unclear. We suspect that the discrepancy between the two results should 
decrease when one propagates the MD in the excited state. Work in this direction
is in progress in our group.

The spectral density can be evaluated as the reweighted
cosine transform of the corresponding bath autocorrelation
function $C\left(t\right)$~\citep{Damjanovic2002,Olbrich2010},
\begin{eqnarray}\label{eqn:spectraldensity}
  J(\omega) &=& \frac{2}{\pi\hbar} \tanh(\beta \hbar \omega/2) \int_0^\infty
  C(t)\cos(\omega t)\ dt.
\end{eqnarray}
With the present data the spectral density exhibits characteristic phonon modes from the
dynamics of the FMO complex, see Fig. \ref{fig:autocorr_spectraldensity} d) first panel. 
However, high-frequency modes tend to be
overpopulated due to the limitation of using classical mechanics.
Most of these modes are the local modes of the pigments, which can be seen from 
the pigment-only calculation in \cite{Ceccarelli2003}.
There are efforts to incorporate quantum effects into the classical
MD simulation in the context of vibrational
coherence~\citep{Egorov1999,Skinner2001,Stock2009}. We are
investigating the possibilities of incorporating corrections based on
a similar approach. Moreover, we also obtain a discrepancy of the
spectral density in the low frequency region. On one hand, 
the origin could lie in the harmonic approximation of the bath modes leading to the $\tanh$
prefactor in Eq. (\ref{eqn:spectraldensity}) or in the force field used in this work.
On the other, the form of the standard spectral density is from \cite{Wendling2000} which measures
fluorescence line-narrowing on a much longer timescale, around ns, than considered in our
simulations (around ps). Assuming correctness of our result, this implies that 
for the simulation of fast exciton dynamics in photosynthetic light-harvesting complexes 
a different spectral density than the widely used one has to be employed.

Site energy cross-correlations between chromophores due to the protein
environment have been postulated to contribute to the long-lived
coherence in photosynthetic systems~\citep{Lee2007}. 
Many studies have explored this issue, e.g. recently 
\cite{Adolphs2006,Rebentrost2009,Nazir2009,Wu2010,Abramavicius2011,Olbrich2011a}.
We tested this argument by de-correlating the site energies.  For each unitary
evolution, the site energies of different molecules at the same time
were taken from different parts of the MD
trajectory. In this way, we could significantly reduce potential cross
correlation between sites while maintaining the autocorrelation
function of each site. As can be seen in Fig. \ref{fig:shuffled_dynamics},  
no noticeable difference between the original and shuffled
dynamics is observed.

\section{Conclusion}
\label{sec:conclusion}
The theoretical and computational studies presented in this article
show that the long-lived quantum coherence in the energy transfer
process of the FMO complex of \emph{Prosthecochloris aestuarii} can be
simulated with the atomistic model of the protein-chromophore complex.
Unlike traditional master equation approaches, we propagate in a 
quantum/classical framework both the system and the environment state 
to establish the connection between the atomistic details of the protein 
complex and the exciton transfer dynamics. 
Our method combines MD simulations and QM/MM with TDDFT/TDA
to produce the time evolution of the excitonic reduced density matrix
as an ensemble average of unitary trajectories.

The conventional assumption of unstructured and uncorrelated site
energy fluctuations is not necessary for our method. No \emph{ad hoc}
parameters were introduced in our formalism. The temperature and
decoherence time were extracted from the site energy fluctuation by
the MD simulation of the protein complex. The simulated dynamics
clearly shows the characteristic quantum wave-like population change
and the long-lived quantum coherence during the energy transfer
process in the biological environment. On this note it is worth 
mentioning that one has to be careful in the choice of force-field
and in the method used to calculated site energies. In fact as 
presented in Olbrich et al.~\cite{Olbrich2011b} a completely different 
energy transfer dynamics was obtained by using the semiempirical ZINDO-S/CIS
to determine site energies.

Moreover, we determined the correlations of the site energy
fluctuations for each site and between sites through the direct
simulation of the protein complex. The spectral density shows the
influence of the characteristic vibrational frequencies of the FMO
complex. This spectral density can be used as an input for quantum
master equations or other many-body approaches to study the effect of
the structured bath.  The calculated linear absorption spectrum we
obtained is comparable to the experimental result, which supports the
validity of our method.  The characteristic beating of exciton
population and pairwise quantum coherence exhibit excellent agreement
with the results obtained by the HEOM
method. It is also worth noting the remarkable agreement of the dephasing
timescales of the MD simulations, the HEOM approach,
and experiment.

Recently, characterization of the bath in the
LH2~\citep{Damjanovic2002,Olbrich2010} and FMO~\citep{Olbrich2011a}
photosynthetic complexes were reported using MD simulation
and quantum chemistry at room temperature. Those studies mostly
focused on energy and spatial correlations across the sites, the linear
absorption spectrum, and spectral density. The detailed study in
~\citep{Olbrich2011a} also suggests that spatial correlations are not relevant
in the FMO dynamics.

This work opens the road to understanding whether biological systems
employed quantum mechanics to enhance their functionality during
evolution.  We are planning to investigate the effects of various factors
on the photosynthetic energy transfer process. These include:
mutation of the protein residues, different chromophore molecules, and
temperature dependence.  Further research in this direction could
elucidate on the design principle of the biological photosynthesis
process by nature, and could be beneficial for the discovery of more efficient
photovoltaic materials and in biomimetics research.

\section*{References}
\bibliography{fmo-md-BiophysJ-Final}{}

\clearpage
\section*{Figure Legends}
\subsection*{Figure~\ref{fig:siteenergy_distribution}}
    Panel {\bf a}: Comparison of the calculated site energies for each BChl molecule
    to the previous works by Wendling et al. and Adolphs et al.~\citep{Wendling2002,Adolphs2006}.
    Our calculation, labeled as MD, was obtained
    using QM/MM calculations with the TDDFT/TDA method at 77K and 300K.
    Vertical bars represent the standard deviation for each site.
    Panel {\bf b}: Marginal distribution of site 1 energy at 77K and 300K. Histograms
    represent the original data, and solid lines correspond to the estimated
    Gaussian distribution.
\subsection*{Figure~\ref{fig:dprates_spectra}}
    Panel {\bf a} shows the calculated dephasing rate for each site 
    at 77K and 300K. Panel {\bf b} shows the simulated linear 
    absorption spectra of the FMO complex at 77K and 300K. 
    They were shifted to be compared to the experimental spectra
    as obtained by Engel through personal communication.
    Panel {\bf c} shows the simulated linear
    dichroism (LD) and circular dichroism (CD) spectra at 77K. Experimental
    spectra were obtained from Wendling et al.\cite{Wendling2002}
    Although TDDFT-calculated spectra shows systematically
    overestimated site energies, the width and overall shape of the spectra
    is in good agreement.
\subsection*{Figure~\ref{fig:population}}
  Panel {\bf a}: Time evolution of the exciton population of each
  chromophore in the FMO complex at 77K and 300K. Panel {\bf b}: Change of
  the pairwise coherence, or concurrence in time. Initial pure states, $\rho_S(0) =
  |1\rangle \langle 1|$ for the top and center panels were propagated
  using the two formulations developed in this article, MD and
  QJC-MD, to utilize the atomistic model of the protein complex
  bath from the MD/TDDFT calculation. Panel {\bf c}  The
  initial state was set to $|6\rangle \langle 6|$ and propagated using the
  MD method.
\subsection*{Figure~\ref{fig:populations_comparison}}
   Comparison of the population dynamics obtained by using the MD method, the
  corrected MD, the hierarchy equation of motion approach and the Haken-Strobl-Reineker
  model at 77K and 300K. Panels on the right correspond to the initial state in site 1 
  and those on the left to an initial state in site 6. 
  All methods show similar short-time dynamics and dephasing, while the long time dynamics
  is different and the different increases as relaxation is incorporated in the various
  methods.  

 \subsection*{Figure~\ref{fig:autocorr_spectraldensity}}
  Panel {\bf a}: Site 1 autocorrelation functions
  using MD and AR(1) processes generated with time constant equal to 2fs, 5fs,
  and 50fs at 77K and 300K.
  Panel {\bf b}: Site 1 population dynamics of MD and AR(1)
  processes with the different time constants at 77K and 300K.
  Panel {\bf c}: The change of pairwise coherence between
  site 1 and 2 of MD and AR(1) processes with the different time
  constants at 77K and 300K.
  Panel {\bf d}: Spectral density of site 1 of the FMO complex from the
  MD simulation at 77K and 300K. They clearly show the
  characteristic vibrational modes of the FMO complex. High-frequency
  modes are overpopulated due to the ultraviolet catastrophe observed
  in classical mechanics. The Ohmic spectral density used by
  Ishizaki and Fleming in~\citep{Ishizaki2009} was presented for
  comparison. The spectral densities of site 1 from AR(1)
  processes are also presented.

\begin{figure}
  \includegraphics[width=3.25in,keepaspectratio]{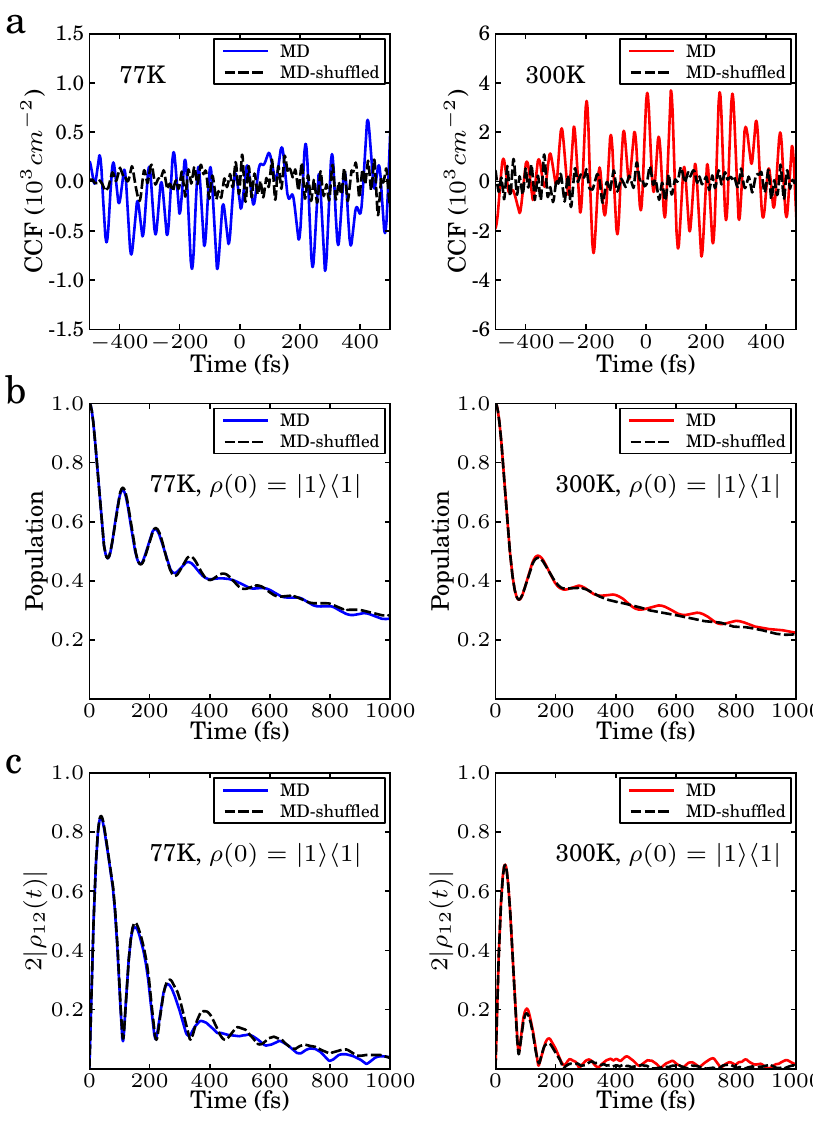}
  \caption{}
  \label{fig:shuffled_dynamics}
\end{figure}

\subsection*{Figure~\ref{fig:shuffled_dynamics}}
  Panel {\bf a}: Cross-correlation function of the original MD trajectory and a
  randomly shuffled trajectory between sites 1 and 2 at 77K and
  300K. Panel {\bf b}: Site 1 population dynamics of the original dynamics
  and the shuffled dynamics at 77K and 300K. {\bf c}, The
  pairwise coherence between sites 1 and 2. Original and shuffled
  dynamics are virtually identical at both temperatures.

\clearpage
\clearpage
\clearpage
\clearpage
\clearpage
\clearpage

\end{document}